\documentclass[12pt]{article}
\usepackage{graphicx}
\usepackage{amsmath, amsthm, amssymb}
\usepackage{hyperref}

\hypersetup{
  breaklinks=true,
  colorlinks=true
}

\newcommand{\veck}{{\boldsymbol k}}
\newcommand{\vecr}{{\boldsymbol r}}
\newcommand{\vecp}{{\boldsymbol p}}
\newcommand{\ii}{{\mathrm i}}
\newcommand{\ee}{{\mathrm e}}
\newcommand{\dd}{{\mathrm d}}

\begin{document}
\title{Diffusive interactions between photons and electrons,\\ an application to cosmology}
\author{L. Marmet\footnote{E-mail: \href{mailto:lmarmet@yorku.ca}{lmarmet@yorku.ca}; ORCID ID \href{https://orcid.org/0000-0001-6087-5480}{0000-0001-6087-5480}}\\ Department of Physics and Astronomy, York University,\\ Toronto Ontario, Canada}
\date{}

\maketitle

\begin{abstract}
The gradient force is the conservative component of many types of forces exerted by light on particles.  When it is derived from a potential, there is no heat transferred to the particle interacting with the light field.  However, most theoretical descriptions of the gradient force use simplified configurations of the light field and particle interactions which overlook small amounts of heating.  It is known that quantum fluctuations contribute to a very small but observable momentum diffusion of atoms and a corresponding increase in their temperature.  This paper examines the contribution to momentum diffusion from a gradient force described as a quantum interaction between electron wave packets and a classical electromagnetic field.  Stimulated transfers of photons between interfering light beams produce a small amount of heating that is difficult to detect in laboratory experiments.  However the solar corona, with its thermal electrons irradiated by an intense electromagnetic field, provides ideal conditions for such a measurement.  Heating from stimulated transfers contributes a large fraction of the observed coronal heating.  Furthermore, the energy removed from the light field produces a wavelength shift of its spectrum as it travels through free electrons.  Theory predicts a stimulated transfer redshift comparable to the redshift of distant objects observed in astronomy.
\end{abstract}

\section{Introduction}
The gradient force manifests itself as a conservative force derived from the gradient of light intensity $\boldsymbol F = \boldsymbol\nabla I$.  A particle interacting with the field of a standing wave will periodically exchange energy and momentum with the intensity pattern generated by interfering light beams.  This gives rise to a force without heat transfer from the radiation to the particle.  Practical applications such as optical trapping and optical tweezers use the conservative property of the force to manipulate atoms without the temperature increase associated with large optical fields.

\begin{figure}[!b]
  \begin{center}
    \includegraphics[width=5.0in]{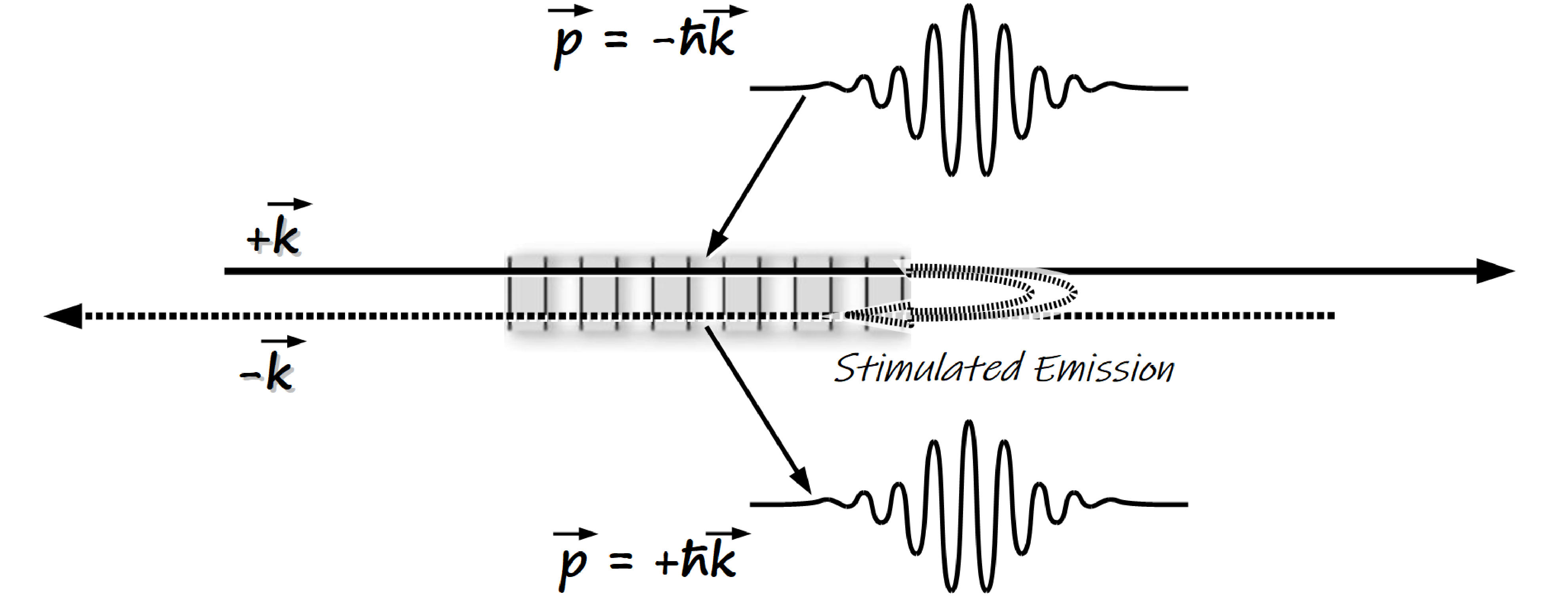}
    \caption{A particle with an initial momentum $\vecp = -\hbar\veck$ interacts with a standing wave.  One photon is absorbed from beam $+\veck$ and is stimulated into beam $-\veck$, resulting in a momentum $+2\hbar\veck$ being transferred to the particle.}
    \label{alm:fig-gradient}
  \end{center}
\end{figure}

At the quantum level, gradient forces are produced by an exchange of momentum between photons and a particle.  For the force to exist, the radiation field must have more than one momentum component,\cite{Fedorov.Letokhov1015.1997} that is, two beams of light must interact simultaneously with a particle.  In a standing wave such as the one depicted in Fig.~\ref{alm:fig-gradient}, photons are removed from one plane wave $\veck$ and stimulated into the counter-propagating wave $\veck' = -\veck$, where $\veck$ is the angular wave vector of the wave.  This \emph{stimulated transfer} changes the momentum of the field by $-2\hbar\veck$ which is taken by the particle as a momentum kick $\Delta\vecp = +2\hbar\veck$.  The force on the particle is then $\boldsymbol F = \dd \vecp/\dd t = \Gamma \Delta\vecp$, where $\Gamma$ is the rate of stimulated transfers.

Because of the quantum nature of the gradient force, each momentum kick from a stimulated transfer increases the width of the momentum distribution of the particles.\cite{Cook976.1980, Gordon.Ashkin1606.1980, Accardi.Lu277.1995}  These quantum fluctuations produce \emph{momentum diffusion} which indicates an increasing temperature of the particles as well as energy being taken away from the radiation field.  In general however, quantum fluctuations of the gradient force are very small compared to the force itself.  Because most experiments use configurations such as standing waves, far detuned excitation, narrow bandwidth lasers, or collimated particles and fields (plane waves) that minimize quantum fluctuations, contributions to momentum diffusion are neglected in most theoretical estimates of the gradient force.

In this paper I give the broad lines of a calculation of the momentum diffusion resulting from the gradient force on electrons\footnote{The gradient force on electrons is also called the \emph{ponderomotive} or \emph{Gaponov-Miller}\cite{Gaponov.Miller242.1958} force.} that includes the small contributions from quantum fluctuations.  The calculation uses these elements:
\begin{itemize}\setlength\itemsep{-4pt}
  \item a statistical mixture of travelling waves instead of standing waves,
  \item a light field with multiple components instead of pure quantum states,
  \item an electron wave packet instead of a plane wave,
  \item momentum-recoil shifts and Doppler shifts,
  \item the density matrix formalism for a large ensemble of particles, and
  \item a second order expansion of the Schr\"odinger equation.
\end{itemize}

With this more complex model of stimulated transfers, a number of effects appear as new properties of the gradient force.  Two of these are examined in details in this paper: the energy transferred to the electron in the form of heat, and the energy lost by light resulting in a spectral shift toward longer wavelengths.

If a frame of reference contains at least two light beams that can interact with the electron, each stimulated transfer gives a momentum kick to the electron initially at rest in that reference frame.  Electrons gain momentum which translates as momentum diffusion and an increase of their temperature.  While the temperature increase would be difficult to detect in the laboratory, it is large enough to be observed in the solar corona.

The second effect is the energy lost by the radiation field as it is transferred to the electrons.  Each stimulated transfer produces a recoil of the electron which Doppler shifts the stimulated photon to a lower energy.  Because every photon transfer is done via stimulated emission, the direction of the beams is preserved while their photons \emph{are being replaced} by photons with a slightly longer wavelength.  The result is a shift of the entire spectrum toward longer wavelengths while preserving its spectral features and directionality.  While this red-shift is difficult to detect in laboratory experiments, the effect is significant for light propagating over large astronomical distances through the free electrons of the intergalactic medium.

The rest of the paper is structured as follows: in Sec.~\ref{alm:sec-stimulated}, I define the framework of the quantum calculation and present the key steps of a calculation of the rate of diffusive stimulated transfers.  Section~\ref{alm:sec-emergent} presents two of the emergent effects of diffusive stimulated transfers that have applications in astrophysics and cosmology.  In Sec.~\ref{alm:sec-discussion}, theoretical results are compared with observations of coronal heating and the astronomical redshift.  I conclude in Sec.~\ref{alm:sec-conclusions} that stimulated transfers play an important role in astrophysical processes.

\section{Stimulated Transfers}\label{alm:sec-stimulated}
The interaction is modelled as an electron interacting with optical plane wave modes
  \[ \boldsymbol{\hat\epsilon}_i E_i \exp \big[ \ii (\veck_i \cdot \vecr - \omega_i t) \big], \]
where $\veck_i$ is the wave vector of wave $i$ with frequency $\omega_i = c|\veck_i|$ and polarization in the direction of the unit vector $\boldsymbol{\hat\epsilon}_i$, and $E_i$ is the electric field amplitude.  In the interaction picture, the sum of the electron's kinetic energy, the energy of the photon field, and an interaction term gives the Hamiltonian
  \[ \tilde H = \frac{\tilde p^2}{2m_e} + \sum_i \hbar\omega_i \, \tilde a_i^\dagger \tilde a_i^{\phantom\dagger}  + \tilde V, \]
where $m_e$ is the mass of the electron, the momentum operator is defined as \mbox{$\tilde\vecp | \vecp \rangle = \vecp | \vecp \rangle$} with the translation property \mbox{$\ee^{+\ii\veck \cdot \vecr} | \vecp \rangle = | \vecp+\hbar\veck \rangle$}, the photon field operator \mbox{$\tilde a | n \rangle = \sqrt{n} | n-1 \rangle$}, $| n; \veck; \vecp \rangle$ denotes an eigenstate vector of the photon number, the photon wave vector, and the electron momentum, respectively, and finally $\tilde V$ is the interaction potential.

The field operator for the transversal component of the vector potential is\cite{CohenTannoudji656.2004}
  \[ \boldsymbol{A} (\vecr) = \sum_i \ii \sqrt{\frac{\hbar}{2\varepsilon_0 \omega_i L^3}} \left[ \tilde a_i \boldsymbol{\hat\epsilon}_i \ee^{\ii \veck_i \cdot \vecr} + \tilde a_i^\dagger \boldsymbol{\hat\epsilon}_i \ee^{-\ii \veck_i \cdot \vecr} \right], \]
where $\varepsilon_0$ is the dielectric permittivity in free space, $\vecr$ is the position vector, and $L$ is the size of the quantization volume according to standard QED procedures in the Coulomb gauge where $\boldsymbol\nabla \cdot \boldsymbol{A} = 0$.

\subsection{Conservative Stimulated Transfers}
Typical derivations of the gradient force use a standing wave represented by a linear superposition $\big| \Psi \big\rangle = \sqrt{\frac{1}{2}} \big( |+\!\veck \rangle -|-\!\veck \rangle \big)$ of single photon states $|\pm\veck\rangle$ produced by photons with momentum $+\hbar\veck$ interfering with retro-reflected photons with momentum $-\hbar\veck$, as depicted in Fig.~\ref{alm:fig-gradient}.

The interaction potential for such a standing wave\cite{Heitler430.1954} is
  \[ \tilde V(t) = \frac{e}{m_e} \boldsymbol A \cdot \vecp = \frac{e}{m_e} \sum_i \ii \sqrt{\frac{\hbar}{2\varepsilon_0 \omega_i L^3}} \left[ \tilde a_i \left( \ee^{+\ii\veck_i \cdot \vecr} - \ee^{-\ii\veck_i \cdot \vecr} \right) + \mbox{H.c.} \right] \boldsymbol{\hat\epsilon}_i \cdot \vecp, \]
where the photon operator $\tilde a$ has an effect on both components of the photon state $\ee^{+\ii\veck \cdot \vecr} - \ee^{-\ii\veck \cdot \vecr}$ simultaneously.  This arises because the wavefunction $\big| \Psi \big\rangle$ is not a statistical mixture of states but an inseparable quantum unit.\cite{Shore.Meystre903.1991}

For long interaction times this potential produces Bragg scattering of a particle by the standing light-wave, where the particle with an initial momentum $\vecp = \hbar\veck$ transfers a photon from the $-\veck$ beam to the $+\veck$ beam via stimulated emission.  After the interaction, the particle's momentum is $\vecp = -\hbar\veck$.  The reverse process is also possible where the particle's momentum is initially $\vecp = -\hbar\veck$ and the interaction transfers a photon from beam $+\veck$ to beam $-\veck$.  This force is derivable from a potential\cite{Gaponov.Miller242.1958} and is practically conservative if we ignore the very small quantum fluctuations.\cite{Cook976.1980, Gordon.Ashkin1606.1980, Javanainen2519.1981}  The energy is maintained over long periods of time, with electrons and the light field exchanging $\pm 2\hbar\veck$ of momentum in a periodic motion described as Pendell\"osung oscillations.\footnote{Named after the pendulum-like motion of the particle.}

\subsection{Diffusive Stimulated Transfers}
This paper focuses on the electron--field interaction for travelling waves.\cite{Shore.Meystre903.1991}  In this case the interaction potential takes the form
  \[ \tilde V(t) = \frac{e}{m_e} \boldsymbol A \cdot \vecp = \frac{e}{m_e} \int \dd\vecp \sum_i \ii \sqrt{\frac{\hbar}{2\varepsilon_0 \omega_i L^3}} \left[ \tilde a_i \ee^{+\ii\veck_i \cdot \vecr} +\tilde a_i^\dagger \ee^{-\ii\veck_i \cdot \vecr} \right] \boldsymbol{\hat\epsilon}_i \cdot \vecp. \]
Here, the main difference is that the photon operator $\tilde a$ has an effect on only one component of the radiation field $\ee^{+\ii\veck \cdot \vecr}$ at a time.

Expanding the Schr\"odinger equation to second order perturbation gives the equation of motion\cite{CohenTannoudji656.2004}
  \[ \Delta \tilde\rho(t) = \frac{1}{i\hbar} \int_{t_0}^{t_0+\Delta t} \dd t \left[ \tilde{V}(t), \tilde\rho(t_0) \right] -\frac{1}{\hbar^2} \int_{t_0}^{t_0+\Delta t} \dd t' \int_{t_0}^{t'} \dd t \left[ \tilde{V}(t'), \left[ \tilde{V}(t), \tilde\rho(t_0) \right] \right], \]
where the density operator in the interaction representation is $\tilde\rho(t) = \ee^{+\ii \tilde{V}t/\hbar} \hat\rho(t) \ee^{-\ii \tilde{V}t/\hbar}$, $\hat\rho(t) \equiv \sum_\chi \bar\rho(t; \chi) | n'; \vecp' \rangle \langle n; \vecp |$, the summation is over $\chi \equiv \{ n', \vecp', n, \vecp \}$, and $\bar\rho(t; \chi)$ are the matrix elements.

The first order term describes a conservative part of the force and can be ignored.  The time evolution of the density matrix is then described by
\begin{equation}
  \Delta \tilde\rho(t)
  = -\frac{1}{\hbar^2} \int_{t_0}^{t_0+\Delta t} \dd t' \int_{t_0}^{t'} \dd t
  \left[ \tilde{V}(t'), \left[ \tilde{V}(t), \tilde\rho(t_0) \right] \right].
  \label{alm:eq-evolution-rho1}
\end{equation}
The first commutator $[ \tilde{V}(t), \tilde\rho(t_0) ]$ describes the annihilation of a photon associated with a momentum transfer $\veck_i$, while the second commutator $[ \tilde{V}(t'), [ ... ] ]$ describes the creation of a photon stimulated with a momentum transfer $\veck_j$.  The net momentum transferred to the electron is $\Delta\vecp = \veck_i - \veck_j$.

To evaluate Eq.~(\ref{alm:eq-evolution-rho1}) we first calculate factors arising from the energy terms.  Without explicit multiplicative constants, the double commutator has this form:
\begin{equation}
  \ee^{+\ii [ \, \omega(\vecp) -\omega(\vecp') \, ] t_0} \ \mathcal{R}_S(\veck_i, \vecp) \ \mathcal{R}_A(\veck_j, \vecp+\hbar\veck_i) \ \big[ \boldsymbol{\hat\epsilon}_i \cdot \vecp \big] \big[ \boldsymbol{\hat\epsilon}_j \cdot (\vecp +\hbar \veck_i) \big] \, \rho(t_0).
  \label{alm:eq-phase}
\end{equation}
These terms represent the time evolution related to kinetic energy, the Doppler and momentum recoils associated with stimulated emission $\mathcal{R}_S$ and annihilation $\mathcal{R}_A$ of a photon, and the energy of the free-electron quiver motion in the electric field, respectively.  Here, $\omega(\vecp) \equiv p^2/(2m_e)$, the momentum recoil term $\mathcal{R}_A(\veck, \vecp) \equiv \exp[+\ii(\xi(\veck, \vecp) -\zeta(\veck))t]$ for the annihilation of a photon, the recoil term for stimulated emission $\mathcal{R}_S(\veck, \vecp) \equiv \exp[-\ii(\xi(\veck, \vecp) +\zeta(\veck))t]$, where the Doppler frequency shift $\xi(\veck, \vecp) \equiv \veck\cdot\vecp/m_e$, the recoil frequency shift
\begin{equation}
  \zeta(\veck) \equiv \hbar k^2/(2m_e),
  \label{alm:eq-recoil-shift}
\end{equation}
and $k \equiv |\veck|$.

Integrating Eq.~(\ref{alm:eq-phase}) over $\vecp$ (from $\tilde{V}(t)$ in the first commutator of Eq.~(\ref{alm:eq-evolution-rho1})) and $\vecp'$ (from $\tilde{V}(t')$ in the second commutator of Eq.~(\ref{alm:eq-evolution-rho1})) gives a phase factor $\Phi \approx 2$.  The approximation $\boldsymbol{\hat\epsilon} \cdot \vecp \simeq \hbar \omega/c = \hbar k$ is used for the free-electron quiver motion at a frequency $\omega$.  Inserting these results in Eq.~(\ref{alm:eq-evolution-rho1}) gives
\begin{align}
  \Delta \rho(t)
  = -\frac{\Phi}{\hbar^2} &\int_{t_0}^{t_0+\Delta t} \dd t' \int_{t_0}^{t'} \dd t \sum_{i, j} \nonumber \\
  &\left[ \frac{e}{m_e} \ii \sqrt{\frac{\hbar}{2\varepsilon_0 \omega_i L^3}} \sqrt{n_i} \hbar k_i \right]
  \left[ \frac{e}{m_e} \ii \sqrt{\frac{\hbar}{2\varepsilon_0 \omega_j L^3}} \sqrt{n_j} \hbar k_j \right] \rho(t_0),
  \label{alm:eq-evolution-rho2}
\end{align}
where contributions from both commutators are displayed explicitly to show that the density matrix evolves at a rate proportional to $\sqrt{n_i}$ related to the electric field of wave $\veck_i$, and $\sqrt{n_j}$ for the electric field of a counter-propagating wave $\veck_j$.

Since $k \approx |\veck_i| \approx |\veck_j|$, Eq.~(\ref{alm:eq-evolution-rho2}) can be simplified to
  \[ \Delta \rho(t) = \frac{\Phi}{\hbar} \left( \frac{e}{m_e} \right)^2 \frac{(\hbar k)^2}{2\varepsilon_0 \omega} \int_{t_0}^{t_0+\Delta t} \dd t' \int_{t_0}^{t'} \dd t \sum_{i, j} \frac{\sqrt{n_i n_j}}{L^3} \rho(t_0). \]

Next we evaluate the time integrals to get a factor $\tau_c \Delta t/2$.  Here $\tau_c$ is a characteristic time on the order of the \emph{collision} time and satisfies $\tau_c \ll \Delta t$, with $\Delta t$ a characteristic time involved in the slow rate of change of $\Delta \rho(t)$.\footnote{See section IV.B.3, p.266 in Ref.~\cite{CohenTannoudji656.2004}.}  Dividing the density matrix $\Delta \rho(t)$ by the coarse time interval $\Delta t$ gives its rate of change
  \[ \frac{\Delta \rho(t)}{\Delta t} = \Phi \frac{e^2}{m_e} \frac{\hbar k^2}{2m_e} \frac{1}{2\varepsilon_0 \omega} \tau_c \sum_{i, j} \frac{\sqrt{n_i n_j}}{L^3} \rho(t_0). \]

Energy conservation appears from the integrals $\int \dd t'$ and $\int^{t'} \dd t$ of Eq.~(\ref{alm:eq-evolution-rho2}), which also imposes a time scale related to the coherence of the scattered waves $\tau_c \approx 1/\omega_{recoil}$.  From Eq.~(\ref{alm:eq-recoil-shift}), the recoil frequency is $\omega_{recoil} = \hbar k^2/(2m_e)$ and the rate of change of the density matrix is
  \[ \frac{\Delta \rho(t)}{\Delta t} = \Phi \frac{e^2}{4\pi\varepsilon_0 m_e c^2} \frac{2\pi c^2}{\omega} \sum_{i, j} \frac{\sqrt{n_i n_j}}{L^3} \rho(t_0). \]

Simplifying the notation with $\lambda \equiv 2\pi/k$, the fine structure constant \mbox{$\alpha = \frac{e^2}{4\pi \varepsilon_0 \hbar c}$}, and the Compton wavelength for the electron \mbox{$\lambda_e = \frac{h}{m_e c}$}, yields
\begin{equation}
  \frac{\Delta \rho(t)}{\Delta t}
  = \frac{\alpha c \lambda_e \lambda}{2} \frac{\Phi}{\pi}
  \sum_{i, \, j}
  \frac{\sqrt{n_i n_j}}{L^3}  \, \rho(t_0).
  \label{alm:eq-rho-evolution}
\end{equation}
This describes the evolution of the density operator as a function of time, which corresponds to the rate of stimulated transfers for an interaction between photons and an electron.

\section{Emergent Effects}\label{alm:sec-emergent}
New effects appear as a result of this detailed calculation of the gradient force.  In addition to the emergent physical effects that appear with stimulated transfers, \emph{Stimulated Transfer heating} and \emph{Stimulated Transfer redshift} produce effects that have direct applications to astrophysics and cosmology.

\subsection{Emergent Physical Effects}

\paragraph{Momentum Diffusion with counter-propagating beams}
An integration over the spatial coordinates $\int \dd\vec{r}$ and $\int \dd\vec{r}\,'$ (implicit in Eq.~(\ref{alm:eq-evolution-rho1})) is required for the evaluation of the density matrix.  This makes Eq.~(\ref{alm:eq-rho-evolution}) correct for nearly counter-propagating beams, which implies that $|\vec{\veck}_i - \vec{\veck}_j| \approxeq 2k$.  However, combinations of beams intersecting at other angles have essentially no contribution to momentum diffusion.

\paragraph{Momentum Diffusion with an electron wave packet and a classical light field}
The electron and field wave functions must contain a continuum of states in order to produce momentum diffusion.  The integrals over the momentum of the electron $\int \dd\vecp \int \dd\vecp'$ and the summations over the photon states $\sum_i \sum_j$ (implicit in an expansion of Eq.~(\ref{alm:eq-evolution-rho1})) will then prevent a simple evolution of the momentum states that would otherwise oscillate between $\vecp = +\hbar\veck$ and $\vecp = -\hbar\veck$ without dissipating any energy.

If the electron wave packet and the photon field are both in a superposition of plane waves that are rich enough to include electron and field quantum numbers both before and after a stimulated transfer,\cite{Fedorov.Letokhov1015.1997} the quantum states will evolve between the states of a continuum and diffuse away from the initial state.

The magnitude of momentum diffusion is obtained from the increase of the electron's momentum from $\vecp = 0$ to $\vecp = \pm 2\hbar \veck$ after a stimulated transfer.  The process convolves an initial momentum distribution of width $\sigma_0$ with the deviation of stimulated transfers $\sigma_{ST} = 2\hbar \veck$ to produce a distribution with a final width $\sigma_\vecp$ that satisfies $\sigma_\vecp^2 = \sigma_0^2 + \sigma_{ST}^2$.  The electrons gain an energy $\Delta E_{ST} = \sigma_{ST}^2/(2m_e) = 2(\hbar k)^2/m_e$ which is characterized by an increase of their temperature.

\paragraph{Rate of Stimulated Transfers}
Setting $n_i = n_j = 1$ in Eq.~(\ref{alm:eq-rho-evolution}) gives the electron density $\sqrt{n_i n_j}\rho(t_0)/L^3 \rightarrow n_e$.  The rate of stimulated transfers \emph{per incident photon} is then
\begin{equation}
  \Gamma(n_e) = \frac{\alpha c \lambda_e \lambda}{2} \frac{\Phi}{\pi} n_e.
  \label{alm:eq-rate}
\end{equation}

\paragraph{Redshift per Stimulated Transfer}
The momentum kick $\hbar|\vec{\veck}_i - \vec{\veck}_j| \approxeq 2\hbar k$ given to the electron produces a recoil frequency-shift $\hbar k^2/(2m_e)$ of the photon.  Each stimulated transfer produces a redshift that is be written as
\begin{equation}
  z_0 = \frac{\Delta\lambda}{\lambda} = \frac{2\hbar k}{m_e c} = 2 \frac{\lambda_e}{\lambda} \simeq \frac{4.8 \times 10^{-12}\mbox{\,m}}{\lambda}.
  \label{alm:eq-z0}
\end{equation}

\subsection{Stimulated Transfer Heating of the Solar Corona}
The high temperature of the solar corona seems counter intuitive, but because radiation losses are very small and thermal conductivity negligible, ``the existence of so high a temperature is not physically impossible.''\cite{Alfven1.1941}
Electrons heated by stimulated transfers absorb energy at a rate $L_h$, causing an increase of the plasma temperature\cite{Brynjolfsson95.2004} until it reaches equilibrium with the power $L_r$ radiated by resonance lines of ionized metals in the solar corona.  At equilibrium $L_h \approx L_r$, which is what is calculated below.

The emission of an optically thin plasma\cite{Rosner.Vaiana643.1978} above the surface of the sun at the radius $R_\odot$ is given by $L_r = n_e(R_\odot) n_H(R_\odot) P(T)$, where $n_e(R_\odot)$ is the electron density, $n_H(R_\odot) \approx n_e(R_\odot)$ is the number density of protons, and $P(T)$ is the radiative loss function at a temperature $T$ for the solar corona.

The measured electron density\cite{Warren.Brooks762.2009} is $n_e(R_\odot) = 2.2 \times 10^{14}$/m$^3$ and the measured temperature of the corona\cite{Morgan.Pickering23.2019} varies from $T_{R\odot} \approx 10^6\mbox{\,K}$ for the quiet sun upwards of $10^7\mbox{\,K}$ for the active sun.

With a typical temperature $T_{R\odot} = 2\mbox{\,MK}$, the radiative loss function of the solar corona is modelled\cite{Rosner.Vaiana643.1978} as $P(T) = 10^{-30.73} T^{-2/3}\mbox{\,Wm$^3$}$, so that $P(T_{R\odot}) = 1.2 \times 10^{-35}\mbox{\,Wm$^3$}$ and the emission of the plasma is
  \[ L_r \approx 0.6 \mu\mbox{W/m$^3$}. \]

Now we turn to the heating rate $L_h$.  Ignoring possible saturation effects, Eq.~(\ref{alm:eq-rate}) for stimulated transfers gives a cross section $\sigma_z =\alpha \lambda_e \lambda \Phi/(2\pi) \approx \, 9 \times 10^{-15}\mbox{\,[m]} \times \lambda$.  The power available to heat electrons in a unit volume is then given by $L_h = \sigma_z z_0 F_e(T) n_e$, where $F_e(T)$ is an \emph{effective} flux density defined below.

Taking the flux density near the solar surface\cite{Cranmer.Winebarger157.2019} equal to $F(R_\odot) = 60$\,MW/m$^2$ and approximating the central wavelength as $\lambda = 550\mbox{\,nm}$, the cross section is $\sigma_z \approx 5.0\,\times\,10^{-21}$~m$^2$ and Eq.~(\ref{alm:eq-z0}) gives $z_0 = 8.7 \times 10^{-6}$.  Electrons \emph{could potentially be heated} at a rate $L_h^* = \sigma_z z_0 F(R_\odot) n_e(R_\odot) \approx 6 \times 10^{-4} \mbox{ W/m$^3$}$ if counter-propagating radiation existed to allow stimulated transfers to occur.

\begin{figure}
  \begin{center}
    \includegraphics[width=5.0in]{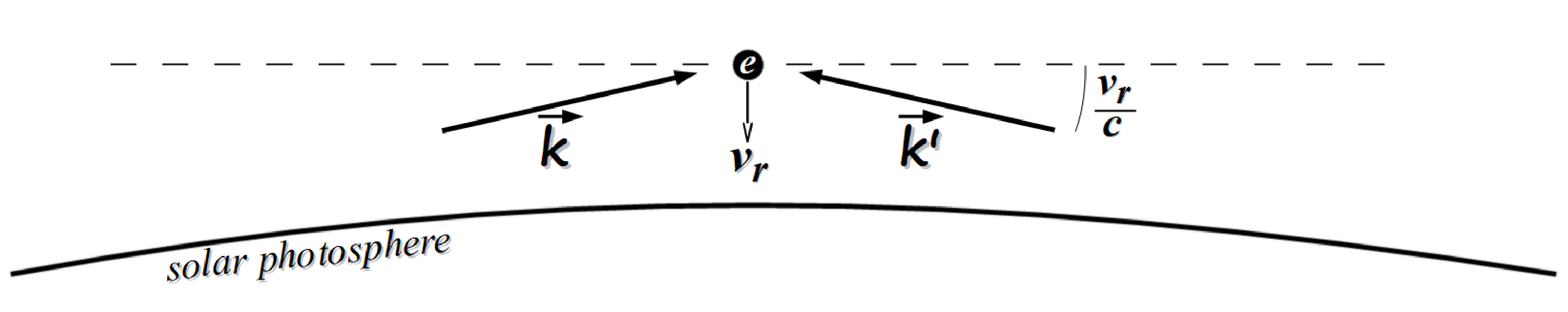}
    \caption{Two light beams $\veck$ and $\veck'$, emitted from opposite sides of the solar limb, interact with an electron (e) moving toward the surface of the sun at the radial velocity $v_r$.  Aberration of light changes their angle by $v_r/c$.  In the reference frame of the electron, the beams satisfy the counter-propagating geometry $\veck_e = -\veck_e'$ required for Stimulated Transfers.}
    \label{alm:fig-aberration}
  \end{center}
\end{figure}

However, radiation coming from the solar surface only covers half of the $4\pi\mbox{\,sr}$ solid angle and has \emph{no matching counterpart coming from space}.  The only counter-propagating light that can interact with an electron near the surface of the sun is produced by aberration of light in the frame of reference of an electron rapidly approaching the sun with a thermal velocity $v_r$, as shown in Fig.~(\ref{alm:fig-aberration}).

The electrons' velocity distribution along the radial direction
  \[ f_v(v_r) = \sqrt{\frac{m_e}{2\pi kT}} \exp \left[ -\frac{m_e v_r^2}{2kT} \right] \]
produces an aberration that lines up pairs of beams over a solid angle $\Omega(v_r) = \pi v_r/c$, with $v_r$ defined as positive toward the sun.  Electrons receding away from the sun are not heated by stimulated transfers.  Integrating the radiance over the solid angle for which the electron ``sees'' counter-propagating beams gives a much smaller effective flux density
  \[ F_e(T) = \frac{F(R_\odot)}{2\pi} \int_0^\infty \Omega(v_r) f_v(v_r) \,\dd v_r = \frac{1}{4}\sqrt{\frac{1}{\pi}}\sqrt{\frac{2kT}{m_ec^2}} F(R_\odot) \] 
that contributes to electron heating.\footnote{In this derivation, the same beams are not counted twice in the calculation of the solid angle and the electrons approaching the sun have a density $n_e/2$.}  At $T_{R\odot} = 2\mbox{\,MK}$, the effective flux density $F_e(T) \approx 0.0052 \times F(R_\odot)$ and the power absorbed by the plasma is
  \[ L_h = \sigma_z z_0 F_e(T) \frac{n_e(R_\odot)}{2} \approx 1.5 \mu\mbox{W/m$^3$}, \]
a value comparable to the radiated power $L_r$ within the approximations of this simplified model.

This shows that in addition to other known coronal heating mechanisms,\cite{Cranmer.Winebarger157.2019} Stimulated Transfer heating contributes sufficient amounts of power to explain the observed high temperatures of the sun's corona.

\subsection{Stimulated Transfer Redshift}
\begin{figure}[!h]
  \begin{center}
    \includegraphics[width=5.0in]{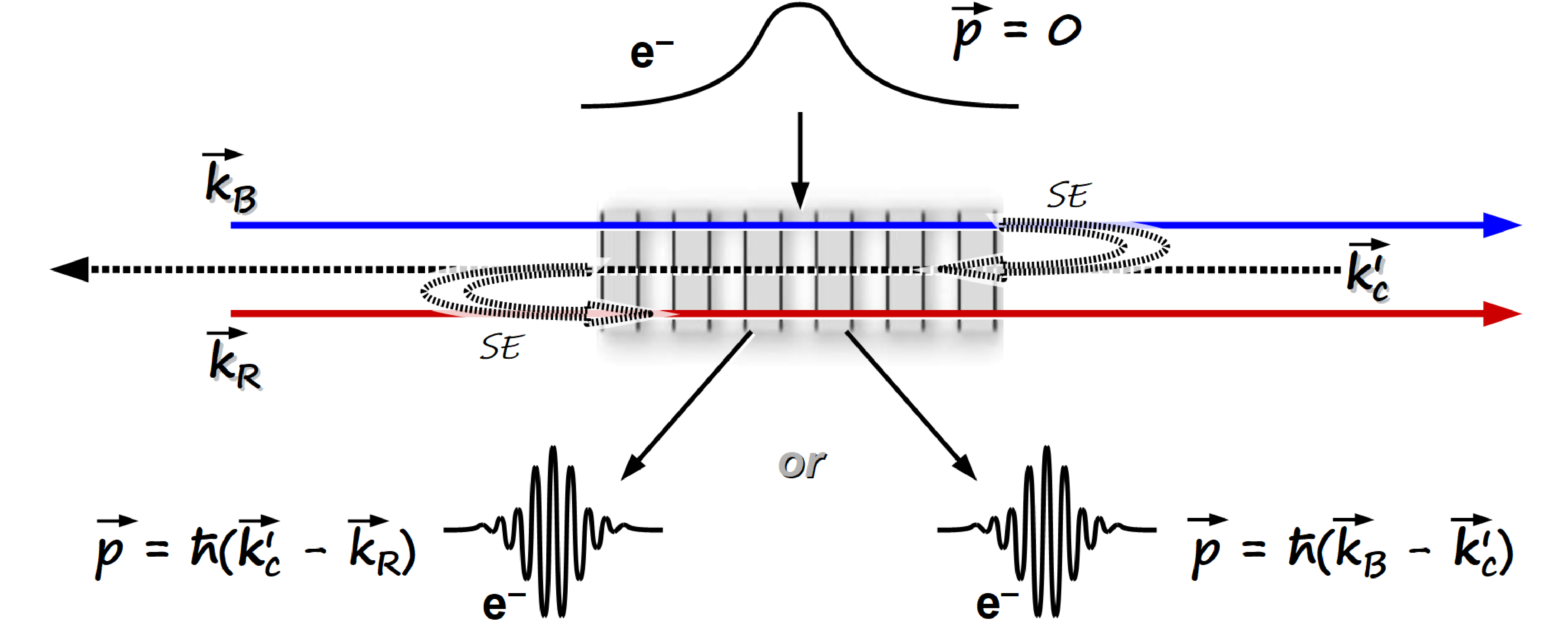}
    \caption{A beam configuration that produces diffusive stimulated transfers.  The electron initially at rest receives a momentum kick from a stimulated transfer and its recoil energy is removed from the light field.  By losing \emph{blue} photons and gaining \emph{red} photons, the spectrum of beam $\veck$ shifts its intensity toward longer wavelengths.  On average, the counter-propagating beam $\veck'$ suffers little changes and the electron is heated by the interactions.}
    \label{alm:fig-stimulated}
  \end{center}
\end{figure}

Light travelling from distant galaxies propagates over large distances through the electrons of the intergalactic medium.  Stimulated transfers along the way will remove energy from the electromagnetic field to produce momentum diffusion of the electrons.  The process can be understood from Fig.~\ref{alm:fig-stimulated}, where a beam $\veck$ coming from the object of interest has two spectral components labelled $\veck_B$ and $\veck_R$ representing a \emph{blue} and a \emph{red} component of its spectrum, respectively.  Another beam $\veck'$ propagates in the opposite direction and is part of the general radiation field in space.

The spectral component $\veck_c'$ participates in two possible interactions.  The first removes a photon from $\veck_B$ and adds a photon via stimulated transfer into $\veck_c'$.  A momentum kick $\Delta\vecp = \hbar(\veck_B - \veck_c')$ is transferred to the electron, which recoils and Doppler shifts the wavelength of the stimulated photon to $\lambda_c' = \lambda_B + 2\lambda_e$.  The second possible interaction with $\veck_c'$ removes a photon from $\veck_c'$ and stimulates it into $\veck_R$.  The momentum kick $\Delta\vecp = \hbar(\veck_c' - \veck_R)$ on the electron Doppler shifts the wavelength of the stimulated photon to $\lambda_R = \lambda_c' + 2\lambda_e$.  The net wavelength shift between $\veck_B$ and $\veck_R$ is $4\lambda_e$, but since either one or the other possibility happens in a transfer event, the wavelength increases on average by $\Delta\lambda = 2\lambda_e$ with each stimulated transfer.  Beam $\veck'$ acts as a catalyst, as it remains mostly unaffected by the stimulated transfers produced by this configuration.\footnote{However, if multiple spectral components are considered for beam $\veck'$, the picture is reversed and it is the spectrum of $\veck'$ that is shifted toward longer wavelengths.}

With a light field rich enough to contain the spectral components $\veck_B$, $\veck_R$, and $\veck_c'$, each stimulated transfer produces a redshift $z_0$ on beam $\veck$.  Because $\veck_B$ and $\veck_R$ can be the spectral components of any beam, stimulated transfers produce a redshift on all radiation travelling through free electrons at a rate $\Gamma(n_e)$.  From  Eq.~(\ref{alm:eq-rate}) and~(\ref{alm:eq-z0}), a wavelength shift of the entire spectrum occurs at the redshift rate
\begin{equation}
  H_z \mbox{\,[s$^{-1}$]} = z_0 \times \Gamma(n_e) = \alpha c \lambda_e^2 \frac{\Phi}{\pi} \, n_e \approx 8.2 \times 10^{-18} \, n_e
  \label{alm:eq-redshift-rate}
\end{equation}
that only depends on the electron density.\footnote{The constant $\alpha c \lambda_e^2$ is $\pi$ times larger than $2hr_e/m_e$ calculated for NTL.\cite{Ashmore53.2006}} Despite the dependence of $z_0$ and $\Gamma(n_e)$ on $\lambda$, the Stimulated Transfer redshift (STz) itself is independent of wavelength and preserves the spectral features of the spectrum at all wavelengths.

Because all energy transfers happen via stimulated emission,\footnote{Of all proposed tired-light mechanisms, only STz and CREIL\cite{Moret-Bailly1215.2003} are based on stimulated emission.} photons added to an already existing light beam acquire the direction of that beam, therefore causing no change of the beam's wavefront properties.  As a result, no blurring occurs and images of distant objects are maintained\cite{Marmet268.2009, Moret-Bailly1215.2003} while the intensity of their spectra is shifted toward longer wavelengths.

\section{Discussion}\label{alm:sec-discussion}
In addition to the known mechanism of coronal heating,\cite{Cranmer.Winebarger157.2019} Stimulated Transfer heating contributes a significant amount of power to the corona.  This is an important piece of the puzzle needed to solve the coronal heating problem that has been described as ``perhaps the longest standing, most frustrating issue yet to be resolved in the solar physics community.''\cite{Gilbert40.2023}

The simple model presented here is unstable to temperature perturbations for $T > 0.25\mbox{\,MK}$ because the modelled radiative loss function decreases with temperature while Stimulated Transfer heating increases with temperature.  To understand precisely how stimulated transfers contribute to the high temperature of the solar corona, a detailed dynamical model will need to include solar limb darkening, energy transport, electron collisions, etc.  However, the above calculation predicts the correct order of magnitude for the energy transfer between photons and electrons due to Stimulated Transfers.  This calculation can therefore be extended with assurance to cosmological processes.

A spectral redshift of radiation propagating through free electrons is produced by Stimulated Transfer redshift (STz).  Stimulated emission causes an energy loss that is a function of the column density of electrons without blurring the images of distant galaxies.  With the simplification that $n_e$ is constant, integrating the energy loss Eq.~(\ref{alm:eq-redshift-rate}) as a function of distance gives $z(D) = \exp \left[ H_z D/c \right] - 1$.  This corresponds to an angular distance
\begin{equation}
  d_A = (c/H_z) \ln(1+z),
  \label{alm:eq-hubble-relation}
\end{equation}
that has the same form as the equation published by Nernst.\cite{Nernst633.1937}  Equating $H_z$ to $H_0 = 73.5\mbox{\,km/s/Mpc}$ taken from Ref.~\cite{Brout.Riess110.2022}, Eq.~(\ref{alm:eq-redshift-rate}) gives $n_e = 0.29\mbox{\,/m$^3$}$, in good agreement\footnote{Of all proposed tired-light mechanisms, only STz and three other models predict the observed electron density in the intergalactic medium, ``NTL,''\cite{Ashmore53.2006} ``Smid's plasma red-shift,'' and ``Bonn's scattering in the intergalactic medium'' (the latter are both described in Ref.~\cite{Marmet55.2018}).} with the measured value $n_e = 0.22_{-0.13}^{+0.11}\mbox{\,/m$^3$}$ of the electron density in the intergalactic medium.\footnote{Based on 21 localized FRBs and a fit to the low redshift values of the Macquart relation shown in Figs.~1 and 2 of Ref.~\cite{Baptista.al57.2024}.  The lowest limit of the distributions shown on these Figures is used to obtain the dispersion measure without the contributions from host galaxies.}

\begin{figure}[!t]
  \begin{center}
    \includegraphics[width=4.5in]{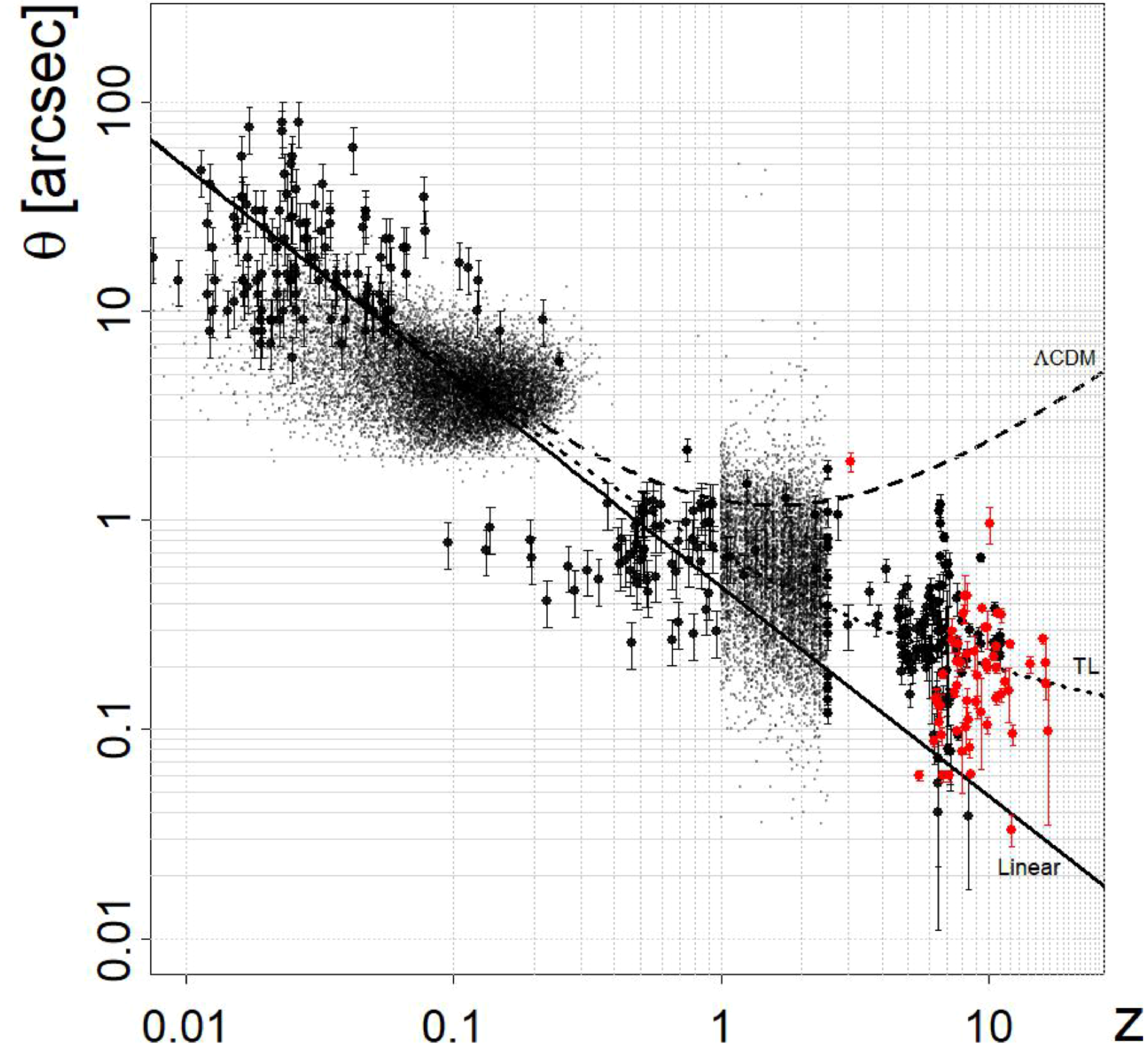}
    \caption{Angular diameters of a 10-kpc-size object as a function of redshift.  Dashed curve: based on $\Lambda$CDM. Dotted curve: based on Eq.~(\ref{alm:eq-hubble-relation}) for a non-expanding cosmology with Tired-Light.  Solid line: based on the Hubble constant $H_0$ and the angular distance as a linear function of $z$.  Measured angular sizes of galaxies from JWST observations (red points) and some pre--JWST observations (black points) are overlapped with the curves.  Figure copied from Lovyagin et al.~(2023).\cite{Lovyagin.Raikov108.2022}}
    \label{alm:fig-angulardiameters}
  \end{center}
\end{figure}

From an observational point of view, not only Eq.~(\ref{alm:eq-hubble-relation}) predicts the linear redshift-distance law found by Hubble\cite{Hubble.Humason43.1931} \emph{for small redshifts}, it also predicts the \emph{Hubble--Humason law}\footnote{In contrast to the Hubble--Lema\^itre law, a redshift-distance relationship named after E.~Hubble and M.~Humason is not the result of space-expansion.} shown in Fig.~\ref{alm:fig-angulardiameters} as measured by JWST\cite{Lovyagin.Raikov108.2022} up to redshifts $z \approx 16$.

To resolve the tensions in cosmology some authors propose a tired-light contribution to the redshift,\cite{Lopez-Corredoira.Marmet37.2022, Gupta7.2024, Shamir703.2024} but the precise mechanism remains unspecified and additional parameters or assumptions are needed to obtain consistency with observations.  By contrast, STz describes a mechanism that predicts astronomical observations from known physics, without any adjustable parameter.

The above derivation contains many simplifications, one of which is an equal number of photons in both the forward beam $\veck$ and the counter-propagating beam $\veck'$.  In real situations, the number of photons differs and $n_i \ne n_j'$ in Eq.~(\ref{alm:eq-rho-evolution}) produces a variable transfer rate as a function of distance from the sources.  As shown in the example of the solar corona, unbalanced values of $n_i$ and $n_j'$ near light sources prevent a large number of stimulated transfers, even if the electron density is higher than the average value in the intergalactic medium.  More detailed modelling is beyond the scope of this paper and will be \mbox{published elsewhere.}

\section{Conclusions}\label{alm:sec-conclusions}
A quantum calculation of momentum diffusion that includes terms usually considered negligible accurately describes momentum diffusion in the gradient force on free electrons.  The photon--electron interaction, calculable from QED, has diffusive properties that remove energy from the light field and increase the temperature of the electrons.  The effect is based on stimulated emission and maintains the directional properties of all light beams.

The calculated heating of electrons in a plasma illuminated by intense light is confirmed by measurements of the solar corona temperature that reaches millions of Kelvins.

Intersecting light beams lose energy to the electrons of the intergalactic medium, resulting in a redshift of their spectral intensity without blurring the images of distant objects.  Photons are replaced by new photons of slightly less energy propagating in the same direction as the original beam.  From the measured electron density in the intergalactic medium, the Stimulated Transfer redshift predicts a redshift--distance relationship that agrees with an observed \emph{Hubble--Humason law} up to $z \approx 16$.

Stimulated transfers play an important role in astrophysical processes and cosmological observations, producing effects that support a very different interpretation of the universe.  In light of all this, ``cautiousness requires not to interpret too dogmatically the observed redshifts as caused by an actual expansion.''\cite{Zwicky802.1935}

\section*{Acknowledgments}
I am especially indebted to Paul Marmet, Chuck Gallo, Jill Delaney, John \mbox{Hartnett}, and Chris Purton for enjoyable and constructive conversations leading to the successful development of the ideas expressed in this paper.  Their invaluable support was instrumental to the progress of this project.  I also thank York University for providing the necessary resources to conduct my research, and \mbox{Prof. Anantharaman Kumarakrishnan} who supported my appointment as an \mbox{Adjunct} Member to the Graduate Program in Physics \& Astronomy.

\bibliographystyle{unsrt}
\bibliography{marmetl-mg17-arxiv}

\end{document}